\documentstyle[preprint,aps]{revtex}
\begin{document}
\draft
\preprint{}
\title{Majorana Neutrinos as the Dark Matters \\ 
in the Cold plus Hot Dark Matter Model}
\author{Noriaki Kitazawa and Nobuchika Okada 
 \thanks{e-mail: n-okada@phys.metro-u.ac.jp}}
\address{Department of Physics, Tokyo Metropolitan University,\\
         Hachioji-shi, Tokyo 192-03, Japan}

\author{Shin Sasaki 
 \thanks{Present address: Department of Physics, Tokyo Metropolitan University, 
Hachioji-shi, Tokyo 192-03, Japan}}
\address{Department of Physics, University of Tokyo, \\
         Bunkyo-ku,  Tokyo 113, Japan}
\preprint{
\parbox{4cm}{
\baselineskip=12pt
TMUP-HEL-9506\\
UTAP-217\\
February, 1996\\
\hspace*{1cm}
}}
\maketitle
\vskip 2cm
\begin{center}
{\large Abstract}
\vskip 0.7cm
\begin{minipage}[t]{14cm}
\baselineskip=19pt
\hskip 4mm
A simple model of the Majorana neutrino with the see-saw mechanism 
is studied, assuming that two light neutrinos are the hot dark matters 
with equal mass of $2.4 {\rm{eV}}$ in the cold plus hot dark
matter model of cosmology.
We find that the heavy neutrino, which is the see-saw partner 
with the remaining one light neutrino, can be the cold dark matter, 
if the light neutrino is exactly massless. 
This cold dark matter neutrino is allowed to have the mass 
of the wide range from $5.9\times 10^2{\rm{eV}}$ 
to $2.2\times 10^7{\rm{eV}}$.   
\end{minipage}
\end{center}
\newpage
\def\beq{\begin{equation}}
\def\dis{\displaystyle}
\def\eeq{\end{equation}}
\def\barr{\begin{eqnarray}}
\def\earr{\end{eqnarray}}
\def\ev{\rm{eV}}
\def\cm{{\rm{cm}}^3}
\def\gev{\rm{GeV}}
Recently Primack {\it{et al.}}   
\cite{primack} pointed out that 
the cold and hot dark matter model agrees very well  
with the observations of the matter distribution in the universe 
with the total density parameter $\Omega = 1$ and 
the Hubble constant 
$h\equiv H_0/100 ~\rm{km} ~\rm{s}^{-1} ~\rm{Mpc}^{-1}=0.5$.
They assumed that two massive neutrinos which 
have nearly degenerate masses $2.4 \ev$ 
play a role of the hot dark matters.  
The hot dark matter, the cold dark matter, and the Baryon 
occupy $20\%$, $72.5\% $, and $7.5\% $ of 
the total density parameter, respectively.

On the other hand, there are some current status for 
the masses and flavor mixings of neutrinos. 
The solar neutrino
deficit \cite{solar}  and the atmospheric neutrino anomaly \cite{atm} 
seem to give the indirect evidences of the  non-vanishing masses 
and flavor mixings of the neutrinos in the view of neutrino 
oscillation. 
In addition,  
recent LSND experiment \cite{lsnd} seems to  
have brought the first direct evidence for neutrino masses 
and flavor mixings in  
the $\overline{\nu_{\mu}}\rightarrow \overline{\nu_{e}}$ oscillation.  
If some species of neutrinos have masses of order eV, they can be appropriate
for the hot components of the dark matter 
in the cold and hot dark matter model. 

In the standard model of the elementary particle physics, 
three species of neutrinos are exactly massless and there is no 
particle which can be the cold dark matter. 
Some extension is needed to include the mass of the neutrinos 
and the cold dark matter.
In this letter, we study the model 
first introduced by Chikashige, Mohapatra and Peccei \cite{cmp}.
This model is a very simple extension of the standard model, 
which includes massive Majorana neutrinos.

We introduce three species of the right-handed 
neutrinos and an electroweak-singlet scalar $\Phi$ as new particles 
to the standard model.
In order to make neutrinos massive, 
two kinds  of the Yukawa couplings are considered. 
The Yukawa interaction is described by 
\barr
{\cal L}_{\rm{Yukawa}}=-g_{{}_Y ij} \overline{\nu _{L i}}\phi \nu _{R j} 
-g_{{}_M ij} \overline{\nu_{R}{}^c_i}\Phi \nu_{R j} +h.c. ,\label{yukawa}
\earr
where $\phi$ is the electric-charge neutral component of 
the Higgs field in the standard model, 
and $i$ and $j$ denote flavors ($i,j=1,2,3$). 
The Dirac and the Majorana mass terms appear by 
the non-zero vacuum expectation values of these scalar fields 
($\langle \phi \rangle \neq 0$ and $\langle \Phi \rangle \neq 0$). 
The mass matrix is given by 
\barr
{\cal L}_{mass}=-\left( \begin{array}{cc}
\overline{\nu_{L}} &  \overline{\nu_{R}{}^c} 
\end{array}\right) 
\left[ \begin{array}{cc}
0 & m_D \\   m_D^T  &  M 
\end{array}\right]
\left( \begin{array}{c}
\nu_{L}{}^c \\ \nu_{R} 
\end{array}\right) + h.c. \label{mat} ,
\earr
where $m_D~(~M~)$ is the 3$\times$3 Dirac (Majorana) mass matrix 
defined by $m_{Dij}=g_{{}_Yij}\langle \phi \rangle$ 
($M_{ij}=g_{{}_Mij}\langle \Phi \rangle$), 
and $\nu_{L}$ and $\nu_{R}$ are the column vectors 
of the three flavors.     
Since the symmetry of the lepton number is spontaneously broken by 
$\langle \Phi \rangle \neq 0$,  
a massless Nambu-Goldstone boson called majoron appears.   
Generally $m_D$ ( $M$ ) is the 3$\times$3 complex (symmetric)  
matrix and we should diagonalize the whole 6$\times $6  mass matrix 
in eq.(\ref{mat}).

As the first approximation, we assume that the off-diagonal elements of the mass 
matrices $m_D$ and $M$ are very smaller than their diagonal elements, 
namely, the effect of the flavor mixing can be neglected. 
By this assumption, we can take $m_D=diag\left[m_1,m_2,m_3 \right]$ and 
$M=diag\left[M_1,M_2,M_3 \right]$. 
Furthermore, we assume the hierarchy between $m_D$ and $M$, namely 
$|m_i/M_i| \ll 1$. 
Then, the see-saw mechanism \cite{seesaw} 
separately works on each generation.  

The six mass eigenstates are described by the weak eigenstates, 
$\nu_L$ and $\nu_R$, as
\barr
\left( \begin{array}{c}
\nu_{\ell} \\
\nu_{h}\end{array}\right)\simeq 
\left[ \begin{array}{cc}
1 & -\epsilon \\ \epsilon & 1 \end{array}\right]
\left( \begin{array}{c}
{\nu}_{L}+{\nu}_{L}{}^c \\ 
{\nu}_{R}+{\nu}_{R}{}^c \end{array}\right) \; , 
\earr
where $\epsilon = diag[{m_1}/{M_1}, {m_2}/{M_2},{m_3}/{M_3}]$, 
and $\nu_{\ell}=(\nu_1^{\ell}\;  \nu_2^{\ell}\; \nu_3^{\ell})^T$ 
and $\nu_{h}=(\nu_1^h \;\nu_2^h \;\nu_3^h)^T $ 
are the light and heavy Majorana fields, respectively. 
The masses for $\nu^{\ell}_i$ and $\nu^h_i$ are 
$m^{\ell}_i\simeq -(m_i)^2  /M_i$ and $m^h_i \simeq M_i$, respectively.

The couplings of the light and heavy neutrinos with the weak bosons are 
given by 
\barr 
{\cal L}_{W\nu}&\simeq &\frac{e}{2\sqrt{2}s} W^{- \mu} \left[
\overline{{\ell}} \gamma_{\mu}(1-\gamma_{5})\nu_{\ell}
+ \overline{{\ell}}\gamma_{\mu}(1-\gamma_{5})\epsilon 
\nu_h\right]+h.c. \; , \label{weak}  \\ 
{\cal L}_{Z\nu}&\simeq &\frac{e}{4s c}Z^{\mu} \left[
\overline{\nu_{\ell}}\gamma_{\mu}(1-\gamma_{5}) \nu_{\ell} + 
\left\{ \overline{\nu_{\ell}} \gamma_{\mu}
(1-\gamma_{5})\epsilon  \nu_h + h.c. \right\} 
+ \overline{\nu_{h}} \gamma_{\mu} (1-\gamma_{5}){\epsilon}^2 \nu_h
\right], \nonumber
\earr
where $\ell $ denotes the column vector of three charged leptons, 
$\ell =(e\; \mu \; \tau)^T $. 
Here we neglect the flavor mixing also in the charged leptons.
The couplings of the heavy neutrinos with the weak bosons are suppressed by 
the small factor $\epsilon $.
These fields $\nu_{\ell}$ and $\nu_h$ also couple  
with the majoron (imaginary part of the field $\Phi$) 
\footnote{
These fields $\nu_{\ell}$ and $\nu_h$ also couple with 
the majoron-Higgs (real part of the field $\Phi $) and the usual Higgs  
in the standard model. However, we do not consider the effects of 
these fields assuming that masses of these couplings assuming that 
these scalar particles are very heavy. 
}.  
According to  eq.(\ref{yukawa}), these couplings are given by 
\barr 
{\cal L}_{\chi \nu}\simeq -\frac{1}{\sqrt{2}}\;  \chi  \; \left[
\overline{\nu_{\ell}}g_{{}_{M}}^{{}_D} \epsilon^2\;  i \gamma_{5}\nu_{\ell} 
-\left\{ \overline{\nu_{\ell}} g_{{}_{M}}^{{}_D} \epsilon \; i \gamma_{5}\nu_h 
+ h.c.\right\}  
+\overline{\nu_h} g_{{}_{M}}^{{}_D}i \gamma_{5}\nu_h
\right]  \label{majoron} \; \; ,
\earr
where the field $\chi$ is the majoron field defined 
by $\chi/\sqrt{2}=\rm{Im}\Phi$, and 
$g_{{}_{M}}^{{}_D}$ is a diagonal matrix,
$g_{{}_{M}}^{{}_D}=diag[g^{{}_{M}}_1,g^{{}_{M}}_2,g^{{}_{M}}_3]
=M/\langle \Phi \rangle$.   
In contrast with eq.(\ref{weak}), the couplings of the light neutrinos 
with the majoron are suppressed by the factor $\epsilon $.

Some parameters are fixed 
according to the  cosmological model 
considered by Primack {\it{et~al}}. 
In their model, the mass spectrum 
of the light neutrinos is constrained as 
$m^{\ell}_1 \ll m^{\ell}_2 \simeq m^{\ell}_3 \simeq 2.4 \ev$.  
Then, the five parameters are left free in our model, 
$g^{{}_M}_1$, $g^{{}_M}_2$, $g^{{}_M}_3$, $m^h_1$ and $m^{\ell}_1$
($\ll 2.4 \ev$).  
Since it is very complicated to analyze  
leaving all of these parameters free, 
we consider the simplest case in this letter.  
We set $g^{{}_M}_1=g^{{}_M}_2=g^{{}_M}_3 \equiv g_{{}_M}$, therefore 
$m^h_1=m^h_2=m^h_3\equiv m_h$.  
Now we have only three free parameters, $g_{{}_M}$, $m_h$, and $m^{\ell}_1$.  
The value of $m^{\ell}_1$ is fixed as the following discussion.

There is a fundamental constraint 
that the particle has to be stable to be the cold dark matter, namely, 
its life time must be longer than the age of the universe 
($t_U \simeq 10^{17}~\rm{s}$).
The heavy neutrinos decay into the light neutrinos and the majoron 
through the coupling in eq.(\ref{majoron}).  
The life time of the $i$-th heavy neutrino is given by 
\beq
\tau_i=\frac{32 \pi}{(g_{{}_M})^2 m^{\ell}_i} \; .
 \label{life}
\eeq
Unless $g_{{}_M}$ is very tiny ($g_{{}_M} < 5.3\times 10^{-15}$),  
two neutrinos, $\nu^h_2$ and $\nu^h_3$, are unstable because of 
$m^{\ell}_2\simeq m^{\ell}_3 \simeq 2.4{\ev}$.  
Since such extremely small coupling constant is unnatural, 
these two neutrinos cannot be naturally the cold dark matters. 
Then, the heavy neutrino $\nu^h_1$ is the only remaining candidate 
for the cold dark matter.   
If we adopt natural value (not so small value) for the Yukawa coupling $g_{{}_M}$, 
the mass of the corresponding light neutrino 
$\nu^{\ell}_1$ must be extremely small, 
for instance, $m^{\ell}_1 < 6.6 \times 10^{-27}\ev$ 
for  $g_{{}_M} \simeq 10^{-2}$. 
Since we cannot believe that such small value of the mass   
is explained by some mechanisms 
(Dirac mass $m_1$ must be extremely smaller than the weak scale), 
we assume that $\nu^{\ell}_1$ is 
exactly  massless by virtue of some symmetries.  
Note that the vanishing $m^{\ell}_1$, or the Dirac mass $m_1=0$, is stable against 
the radiative correction, 
since the $U(1)$ symmetry of the phase rotation of the fields $\nu_{1L}$ 
and $e$ forbids the generation of the Dirac mass $m_1$. 
Furthermore, it should be stressed that 
only the mass difference is important for the neutrino oscillation phenomena, 
but not the absolute value of the mass.   
There is nothing wrong with the massless $\nu_1^{\ell}$. 
According to this assumption, 
the heavy neutrino $\nu_1^h$ can be stable cold dark matter.
In the following discussion, our aim is to investigate 
the cosmologically allowed region of the  mass $m_h$ 
and the coupling constant $g_{{}_M}$. 

Let us consider the dynamical evolution of the early universe 
in this model.
At the very high temperature, we assume that 
all the particles are in thermal equilibrium.  
As the temperature cools down, some particles decouple from the 
thermal equilibrium at each specific temperature called decoupling temperature. 
It is convenient for considering the evolution of the universe 
to separate the matter contents into two parts. 
One is the  `heavy neutrino-majoron system' which includes 
three heavy neutrinos and the majoron. 
The other is the `electroweak system' in which all the other particles are included. 
These two systems weakly interact with each other 
through the couplings suppressed by the see-saw factor $\epsilon $
in eqs.(\ref{weak}) and (\ref{majoron}).  
We assume that at the temperature $T^{EW}_D$ the `heavy neutrino-majoron system' 
decouples from the `electroweak system', and 
that the particles in the `heavy neutrino-majoron system' are still in 
thermal equilibrium after this decoupling. 
Therefore, the another decoupling temperature $T^{\chi}_D$ 
is lower than $T^{EW}_D$. 
Here $T^{\chi}_D$ is the temperature at which 
the heavy neutrinos and the majoron no longer interact with each other.

Note that the temperatures of these two systems are different after the 
decoupling at $T^{EW}_D$. 
The temperature of the heavy neutrino as the cold dark matter 
in the present universe $T_{CDM}$ is different from the 
temperature of photon at present, $T_r=2.7\rm{K}$. 
The reheating factor $\alpha_R$ is estimated by considering 
the reheating of photon caused by  
the charged particle and the anti-particle annihilation.  
In addition to this usual factor $\alpha_R$, the cooling factor $R$ should 
be introduced in our model in order to include cooling effect by 
the decaying unstable neutrinos. 
As soon as the `heavy neutrino-majoron system' 
 decouple from the `electroweak system' at $T_D^{EW}$,  
two unstable neutrinos $\nu_2^{\ell}$ and $\nu_3^{\ell}$ 
decay into the light neutrinos and the majoron.  
Then, the `heavy neutrino-majoron system' cools down 
and the `electroweak system' is heated up, because the energy of 
the `heavy neutrino-majoron system' flows to the `electroweak system' by 
the emission of the light neutrinos.  
Since the degrees of freedom of the `electroweak system' is far 
larger than that of the `heavy neutrino-majoron system', 
we can ignore the heating effect of the `electroweak system'. 
The cooling factor $R$ is defined by 
\barr
R= \frac{T_D^{\chi}}{T_{EW}^{\chi}}\; , \label{Rdeff} 
\earr
where $T_{EW}^{\chi}$ is the temperature of the `electroweak system' 
when the heavy neutrinos decouple from the majoron. 
Since $T_D^{\chi}<T^{EW}_D$ as assumed above, 
the factor $R$ is less than unity.  
The temperature $T_{CDM}$ is written as 
\barr
T_{CDM} = \alpha_R \; R \; T_r \; . \label{tcdm}
\earr
by using the factors $\alpha_R$ and $R$. 

Now we can write down the condition for the heavy neutrino $\nu^h_1$ 
to be the cold dark matter, 
$\rho_{\nu^h_1}=\rho_{CDM}$. 
Here $\rho_{CDM}=1.9\times 10^{-6} {\rm{GeV}}/{\rm{cm}}^3$ is 
the energy density of the cold dark matter in the present universe 
based on the model considered by Primack {\it{et~al}}.   
The present energy density of the cold dark matter neutrino $\nu^h_1$  
is given by 
\barr 
\rho_{\nu^h_1}=m_h \; n(T_D^{\chi})\; 
 \left(\frac{T_{CDM}}{T_D^{\chi}}\right)^3 \label{rhocdm},
\earr
where $n(T_D^{\chi})$ is the number density of the cold dark matter at the 
decoupling temperature $T_D^{\chi}$.  
This number density is given by 
\barr
n(T_D^{\chi})&=&\frac{1}{\pi^2}\int_{0}^{\infty} 
\frac{p^2 dp}{\exp[\frac{E}{T_D^{\chi}}]+1}  \label{deffx} \\ \nonumber
&=& f(x_D) \; (T_D^{\chi})^3 ,  
\earr
where $E=\sqrt{p^2+m_h^2}$, $x_D = T_D^{\chi}/m_h$, and $f(x_D)$ is defined by 
\barr
f(x_D)=\frac{1}{\pi^2}\int_{0}^{\infty}
\frac{y^2 dy}{e^{\sqrt{y^2+x_D^{-2}}}+1}\; .
\earr
Considering that the cold dark matter should decouple in the 
non-relativistic regime, we provide 
the upper bound of $x_D$ as $x_D \leq 1$.  
By using the condition, $\rho_{\nu^h_1}= \rho_{CDM}$, and  
eqs.(\ref{tcdm})-(\ref{deffx}), 
the mass $m_h$ is given by  
\barr
m_h(x_D,R)= 1.1 \times 10^{-9}\;\alpha_R^{-3}\; f(x_D)^{-1} \;  R^{-3} 
\; \; {\rm{GeV}}
\label{mass}
\earr
as the function of $x_D$ and $R$. 
 
On the other hand, the coupling constant $g_{{}_M}$ is obtained from  
the definition of the decoupling temperature $T_D^{\chi}$. 
The decoupling temperature is defined by \cite{kt}
\barr
n(T_D^{\chi})\; \langle \sigma |v| {\rangle}_{T_D^{\chi}}  = H \; , \label{dt}
\earr
where $\langle \sigma |v|{\rangle}_{T_D^{\chi}} $ is the average value of the 
annihilation cross section of a heavy neutrino times relative velocity, 
and $H$ is the Hubble parameter.  
For the non-relativistic heavy neutrino, 
we obtain 
\barr
\langle \sigma |v|{\rangle}_{T_D^{\chi}} = \frac{(g_{{}_M})^4}{128 \pi}\;  
\frac{T_D^{\chi}}{(m_h)^3}
\; ,  \label{cross}
\earr
by considering the heavy neutrino annihilation process,
$\nu_1^h \nu_1^h \rightarrow \chi\chi$. 
The  Hubble parameter $H$ is given by
\barr
H = \left(\frac{8\pi^3}{90}\right)^{\frac{1}{2}}\sqrt{g_*}
\frac{T^2}{M_P}\;  ,
\earr
where $ M_P\simeq 1.2 \times 10^{19}\gev$ is the Planck mass, 
and $g_*$ is the  total degrees of freedom 
of the all particles in thermal equilibrium. 
Note that $T$ is not equal to  $T_D^{\chi}$. 
Since the degrees of freedom of the `electroweak system' are far larger 
than that of the `heavy neutrino-majoron system', 
the expansion rate of the universe, or the Hubble parameter, 
is approximately controlled only by the `electroweak system'. 
Therefore, we can set $T$ and $g_*$ the temperature $T_{EW}^{\chi}$ 
and the degrees of freedom of the `electroweak system', respectively.    
From the definition of $R$ in eq.(\ref{Rdeff}), the Hubble parameter is 
rewritten as 
\barr
H = \left(\frac{8\pi^3}{90}\right)^{\frac{1}{2}}\sqrt{g_*}
\frac{(T_D^{\chi})^2}{M_P R^2} \label{hubble} \; .
\earr
Substituting eqs.(\ref{deffx}), (\ref{cross}) and (\ref{hubble}) 
into eq.(\ref{dt}) and eliminating $m_h$ by using eq.(\ref{mass}), 
$g_{{}_M}$ is given by 
\barr
g_{{}_M}(x_D,R)=5.0\times10^{-7}\; g_{*}^{1/8} \; \alpha_R^{-3/4}\; 
x_D^{-1/2}\; f(x_D)^{-1/2}  \; R^{-5/4}\; .  \label{gm}
\earr
as the function of $x_D$ and $R$.
Since we obtain both $m_h$ and $g_{{}_M}$ as the functions of $x_D$ and $R$, 
one line is drawn in the $m_h$-$g_{{}_{M}}$ plane 
for one fixed values of $R (\leq 1)$ varying $x_D$ from zero to unity.    
The allowed region of $m_h$ and $g_{{}_M}$ is very large, if the value of 
$R$ is absolutely free. 

Next we estimate the cooling factor $R$  
by using the `sudden-decay' approximation 
for two unstable heavy neutrinos $\nu^h_2$ and  $\nu^h_3$. 
We approximately consider that all the unstable neutrinos decay and 
disappear at once 
when the age of the universe is equal to their life time, 
$\tau$ $(=\tau_2 \simeq\tau_3)$ 
\footnote{
The unstable neutrinos $\nu_2^h$ and  $\nu_3^h$ do not start decaying 
from $t=0$, but $t=t_D^{EW}$ at which 
they decouple from the `electroweak system'.  
However, we can approximately set $t_D^{EW}=0$, 
because $t_D^{EW} \ll \tau$ is satisfied in our final result.
}.
In addition to this approximation, we assume that the disappeared 
$\nu_2^h$ and  $\nu_3^h$ are quickly supplied by the majoron annihilation, 
and the thermal equilibrium is recovered.    
The same situation is expected to occur also 
at the age $t=2 \tau, 3\tau$ and so on,  
until the temperature of the `heavy neutrino-majoron system' cools down to the 
decoupling temperature $T_D^{\chi}$.  
According to these approximations, the ratio $\tilde{T}_{\chi}/T_{\chi}$ 
can be estimated, 
where $\tilde{T_{\chi}}$ is the temperature   
of the `heavy neutrino-majoron system'  
just after quick supplement and $T_{\chi}$ 
is the one just before the `sudden-decay'. 

The energy density of the `heavy neutrino-majoron system', 
$\rho_{hm}$, is described in two different ways. 
Since the heavy neutrinos and the majoron are in thermal equilibrium  
just after quick supplement, we obtain 
\barr
\rho_{hm} = \frac{\pi^2}{30} (1+\frac{7}{4} \times 3)\; \tilde{T}_{\chi}^4 
\; . 
\earr
On the other hand, just after `sudden-decay' (before quick supplement), 
$\rho_{hm}$ is given by 
\barr
\rho_{hm} = \frac{\pi^2}{30} (1+\frac{7}{4} \times 3)\; T^4_{\chi}  
-\frac{\pi^2}{30} (\frac{7}{4}\times 2 \times \frac{1}{2})
\; T^4_{\chi} \; . 
\earr
The second term denotes the loss of the energy density due to the 
emission of the light neutrinos. 
From these two expressions of $\rho_{hm}$,
we can obtain $\tilde{T}_{\chi}/{T}_{\chi}= (18/25)^{1/4}$. 
Since the `sudden-decay' and the quick supplement recurrently occur, 
we obtain ${T_{hm}}/T_{EW}=(18/25)^{{\rm{n}}/4}$
at the age $t={\rm{n}} \tau$, 
where $T_{hm}$ and $T_{EW}$ are the temperature of the 
`heavy neutrino-majoron system' and the `electroweak system' 
at the same age, respectively, 
and ${\rm{n}}$ is the positive integer. 
This ratio is translated to the smooth function of the age of the universe $t$: 
${T_{hm}}/T_{EW}=(18/25)^{t/{4\tau}}$.  
Therefore, the cooling factor $R$ is given by 
\barr
R=\left(\frac{18}{25}\right)^{\frac{t_D^{\chi}}{4\tau}}\; , \label{Rapp}
\earr
where $t_D^{\chi}$ is the age of the universe at which 
the heavy neutrino decouple from the majoron.

On the other hand, the definition of $R$ in eq.(\ref{Rdeff}) is rewritten as  
\barr
R=\frac{T_D^{\chi}}{T_{EW}^{\chi}}=\frac{m_h x_D}{T_{EW}^{\chi}} \; . \label{Rdef}
\earr
Since $T_{EW}^{\chi}$ is described by $t_D^{\chi}$ by using 
the relation between the Hubble parameter 
and the age of the universe:
\barr 
H=\left(\frac{8\pi^3}{90}\right)^{\frac{1}{2}}\sqrt{g_*}
\frac{(T_{EW})^2}{M_P} =\frac{1}{2t} \; , 
\earr
we obtain 
\barr
R=m_h x_D \left(\frac{90}{8\pi^3 g_{*}}\right)^{-\frac{1}{4}}
\sqrt{\frac{2t_D^{\chi}}{M_P}}\; . \label{Rt}
\earr

By using eqs.(\ref{life}), (\ref{Rdef}), and (\ref{Rt}), 
we can obtain a relation among $m_h$, $g_{{}_M}$,  $x_D$, and  $t_D^{\chi}$ 
as 
\barr
\left(\frac{18}{25}\right)^{\frac{t_D^{\chi}}{4\tau(g_{{}_M})}}
=\frac{m_h x_D}{T_{EW}^{\chi} (t_D^{\chi})} \; .  \label{eq3}
\earr
Now we obtain three independent relations, 
eqs.(\ref{mass}), (\ref{gm}), and (\ref{eq3}) 
for $m_h$, $g_{{}_M}$,  $x_D$ and  $t_D^{\chi}$.  
Therefore, we can draw one line in the $m_h$-$g_{{}_M}$ plane.  
The result of numerical calculations for these relations is shown in Fig.1 
(upper solid line for various values of $x_D\leq 1$).   

However, note that our approximations underestimate the value of $R$. 
Because the correct amount of the decaying $\nu_2^h$ and $\nu_3^h$ 
is clearly smaller than that estimated by `sudden-decay' approximation. 
Furthermore we provide next decaying neutrinos
by quickly supplement approximation, 
although the disappeared $\nu_2^h$ and $\nu_3^h$ are not so 
quickly supplied. 
Since both $m_h$ and $g_{{}_M}$ are decreasing functions of $R$ 
as can be seen in eqs.(\ref{mass}) and (\ref{gm}), 
the upper solid line in Fig.1 is interpreted as 
the upper bound of the allowed region. 

There exists the upper bound of $m_h$, 
since we assumed that all particles are in thermal equilibrium at 
very high temperature.  
Then there should exist a value of temperature $T (\neq 0)$ 
which can satisfy the condition as follows: 
\barr
\frac{n\; \langle \sigma |v| \rangle} {H} \geq 1 \; . \label{cond}
\earr
Here $n$ is the number density of the heavy neutrino and 
$\langle \sigma |v| \rangle$ is the average value of 
the annihilation cross section of the heavy neutrino 
times relative velocity.  
Considering the annihilation processes 
$\nu_h \nu_h \rightarrow \nu_{\ell}\nu_{\ell}$, 
$\ell \overline{\ell}$, or $q\overline{q}$ ($q$ denotes quark) 
according to the weak 
interaction of eq.(\ref{weak}), we obtain 
\barr
n\; \langle \sigma |v| \rangle =
\frac{4 N G_F}{3\pi^2} \left(\frac{m_{\ell}}{m_h}\right)^2
\int_0^{\infty}\frac{p^4 dp}{e^{E/T}+1}
\frac{M_Z^4}{(4E^2-M_Z^2)^2+M_Z^2 \Gamma_Z^2}\; , \label{sigweak}
\earr
where $E=\sqrt{p^2+m_h^2}$, $m_{\ell} \simeq2.4 \rm{eV}$,   
$G_F\simeq 1.17 \times 10^{-5} \gev^{-2}$ is the Fermi constant,    
$M_Z$ is the mass of the $Z$ boson, 
and $\Gamma_Z=2.5 \gev$ is the total decay width of 
the $Z$ boson 
\footnote{
The annihilation cross section 
intermediated by the majoron is far smaller than that intermediated 
by the gauge boson 
in the region of $m_h$ and $g_{{}_M}$ of our final result.
Then, it can be ignored.  
}.
The factor $N$ is defined by 
$N=(I_3-Q \sin^2 \theta_W)^2 + (I_3)^2$, where $I_3$ and $Q$ 
are the third component of the weak isospin ($\pm \frac{1}{2}$) 
and the electric charge 
of the final state fermion, respectively.  
Summing $N$ for all the possible final state fermions, 
we obtain $N\simeq 7.3$. 
By numerical calculation of eqs.(\ref{cond}) and (\ref{sigweak}), 
we obtain the upper bound $m_h \leq 22 \; \rm{MeV}$.

Finally, we consider the constraint 
from the big bang nucleosynthesis (BBN).  
The number of species of the light neutrinos is constrained as 
$N_{\nu} \leq 3.04$ \cite{bbn2}.  
The contribution of new particles (three heavy neutrinos and 
the majoron) to the energy density in the BBN era ($\simeq 1 \rm{MeV}$)
have to be small enough in comparison with 
\barr
\rho_{\Delta N}=\frac{\pi^2}{30}\; 
 \frac{7}{4} \; \Delta N \times (1\rm{MeV})^4 \; ,
\earr
where $\Delta N={\rm{max}}(N_{\nu})-3=0.04$. 
The energy density of the new particles is given by 
\barr
\rho_{new}=\frac{\pi^2}{30}\;  (1+\frac{7}{4}\times 3) \;
\times (\tilde{\alpha_R}R)^4\; (1\rm{MeV})^4 \; ,  
\earr
where $\tilde{\alpha_R}$ is the ordinary reheating factor 
at the BBN era, 
$\tilde{\alpha_R}= 
\left(g_{*}(1{\rm{MeV}})/g_{*}(T_D^{EW}) \right)^{1/3}$.  
Therefore, we can obtain the upper 
bound of $R$:  
\barr
R \leq \left(\frac{7}{25}\times \Delta N  \right)^{\frac{1}{4}} {\tilde{\alpha_R}}^{-1}
\; . \label{Rub}
\earr
This provides the lower bond on the allowed region of $m_h$ and $g_{{}_M}$, 
since both $m_h$ and $g_{{}_M}$ are the decreasing functions of $R$.  
The bound 
\footnote
{
If we refer the more loose bound, $N_{\nu} \leq 3.3$ \cite{bbn1},
the upper bound of $R$ becomes little larger
and the allowed region becomes a little larger.
}
is shown in Fig.1 as the 
lower solid line for various values of $x_D$.   

Our final result for the mass of the cold dark matter and 
the coupling to the majoron
is shown in Fig.1.
The region among the dashed line, the upper solid line, the lower 
solid line,  
and the horizontal line of the upper bound on $m_h$ 
is allowed in our analysis.  
The allowed region of $m_h$ and $g_{{}_M}$  covers over 
about five and three orders of magnitude, respectively. 
The mass matrices of the see-saw type are realized in this 
allowed region of $m_h$. 
The right hand side of the dashed line in Fig.1 satisfies 
the constraint that the cold dark matter should decouple from the 
majoron in non-relativistic regime. 
Requiring the fact 
that the heavy neutrino has been in thermal equilibrium of 
the `electroweak system' once, 
the upper bound on $m_h$ $(\leq 22{\rm{MeV}})$ is obtained. 
The `sudden-decay' and `quick supplement' approximations provide  
the upper bound of the allowed region (upper solid line in Fig.1).   
The lower bound of the allowed region is obtained 
by the BBN constraint (lower solid line in Fig.1).  

Here we would like to mention the constraint from the characteristic 
mass scale of the free streaming of the cold dark matter.  
The free streaming length is roughly estimated as \cite{kt} 
\barr  
\lambda_{FS} \simeq \left(\frac{1{\rm{keV}}}{m_h}\right) \;  
\left(\frac{T_{CDM}}{T_r}\right)   \; {\rm{Mpc}}
= \left(\frac{1{\rm{keV}}}{m_h}\right) \alpha_R R \; {\rm{Mpc}}\; ,
\earr
and we obtain the characteristic mass scale of the free streaming as  
\barr
M_{FS}= \frac{4}{3}\pi \lambda_{FS}^3 \; \rho_{_{CDM}} 
\simeq2.1\times 10^{20} M_{\odot} 
\left(\frac{\alpha_R R}{m_h({\rm{eV}})}\right)^3 \label{mfs}\; , \nonumber
\earr
where $M_{\odot}$ is the solar mass. 
This scale means the lower limit of the scale of structure 
which can be formed by the effect of the cold dark matter. 
If we consider that the scale of the globular clusters ($10^6M_{\odot}$) 
should be explained by the cold dark matter, 
more strict lower bound on $m_h$ is obtained. 
From the condition of $M_{FS}\leq 10^6M_{\odot}$ and 
using eqs.(\ref{mass}) and (\ref{gm}), 
we obtain the bound $m_h > 6.3\times 10^3 {\rm{eV}}$.  

In conclusion, 
we studied whether the heavy Majorana neutrino can be the cold dark matter
or not in the cold plus hot dark matter model 
considered by Primack et al.  
The model of the Majorana neutrino first introduced 
by Chikashige, Mohapatra, and Peccei was considered 
as the simple extension of the standard model.
We found that if a light neutrino is exactly massless, 
the heavy neutrino, which is the see-saw partner of the massless neutrino, 
can be the cold dark matter, provided that 
other two light neutrinos play the role of the hot dark matter. 
Therefore, both the hot and cold dark matters are Majorana neutrinos. 
We obtained the wide allowed region in the $m_h$-$g_{{}_M}$ plane 
by considering the cosmological arguments.   

\begin{figure}
\caption{The allowed region of the mass of 
 the cold dark matter $m_h$ and the coupling 
 to the majoron $g_{{}_M}$. The dashed line is the line of $x_D=1$ 
  for various values of $R$. 
  The upper solid line is the upper bound for the allowed region. 
  The dots on this line correspond to $x_D=1,0.6,0.4,0.2$ from below, respectively. 
  The lower solid line is the lower bound for the allowed region. 
   The dots on this line correspond to $x_D=1, 0.6,0.4,0.2, 0.1$ 
   from below, respectively.      
   The dotted horizontal line is the upper bound of the mass, 
   $m\leq 22\rm{MeV}$. 
   The region among these four lines is allowed in our analysis.}
\end{figure}

\begin{references}
\bibitem{primack}  
J. R. Primack, J. Holtzman, A. Klypin and D. O. Caldwell,  
Phys. Rev. Lett. 74 (1995) 2160.
\bibitem{solar}  
B. T. Cleveland et al., Nucl. Phys. B (Proc. Suppl.) 38 (1995) 47; 
K. S. Hirata et al., Phys. Rev. D 44 (1991) 2241; 
Y. Suzuki, Nucl. Phys. B (Proc. Suppl.) 38 (1995) 54; 
A. I. Abazov et al., Phys. Rev. Lett. 67 (1991) 3332; 
J. N. Abdurashitov et al., Nucl. Phys. B (Proc. Suppl.) 38 (1995) 60; 
P. Anselmann et al., Phys. Lett. B 285 (1992) 376; 
ibid 327 (1994) 377. 
\bibitem{atm}
K. S. Hirata et al., Phys. Lett. B 205 (1988) 416; 
ibid B 280 (1992) 146.  
\bibitem{lsnd}
C. Athanassopoulos et al., Phys. Rev. Lett. 75 (1995) 2650; 
J. E. Hill, ibid 75 (1995) 2654.
\bibitem{cmp} 
Y. Chikashige, R. N. Mohapatra and R. D. Peccei, 
Phys. Lett. B 98 (1981) 265.  
\bibitem{seesaw}
T. Yanagida, 
in: Proc. of the Workshop on the Unified Theory and Baryon 
Number in the Universe, eds. O. Sawada and A. Sugamoto 
(KEK report 79-18, 1979), p. 95;   
M. Gell-Mann, P. Ramond and R. Slansky, 
in: Supergravity, eds. P. van Nieuwenhuizen and D. Z. Freedman  
(North Holland, Amsterdam, 1979), p. 315. 
\bibitem{kt}
E. W. Kolb and M. S. Turner, The Early Universe 
(Addison-Wesley Publishing Co., California, 1990).  
\bibitem{bbn2}
P. Kernan and L. Krauss, Phys. Rev. Lett. 72 (1994) 3309.  
\bibitem{bbn1}
T. P. Walker, G. Steigmann, D. N. Olive and H. Kang, 
Astrophys. J. 376 (1991) 51.  
\end{references}
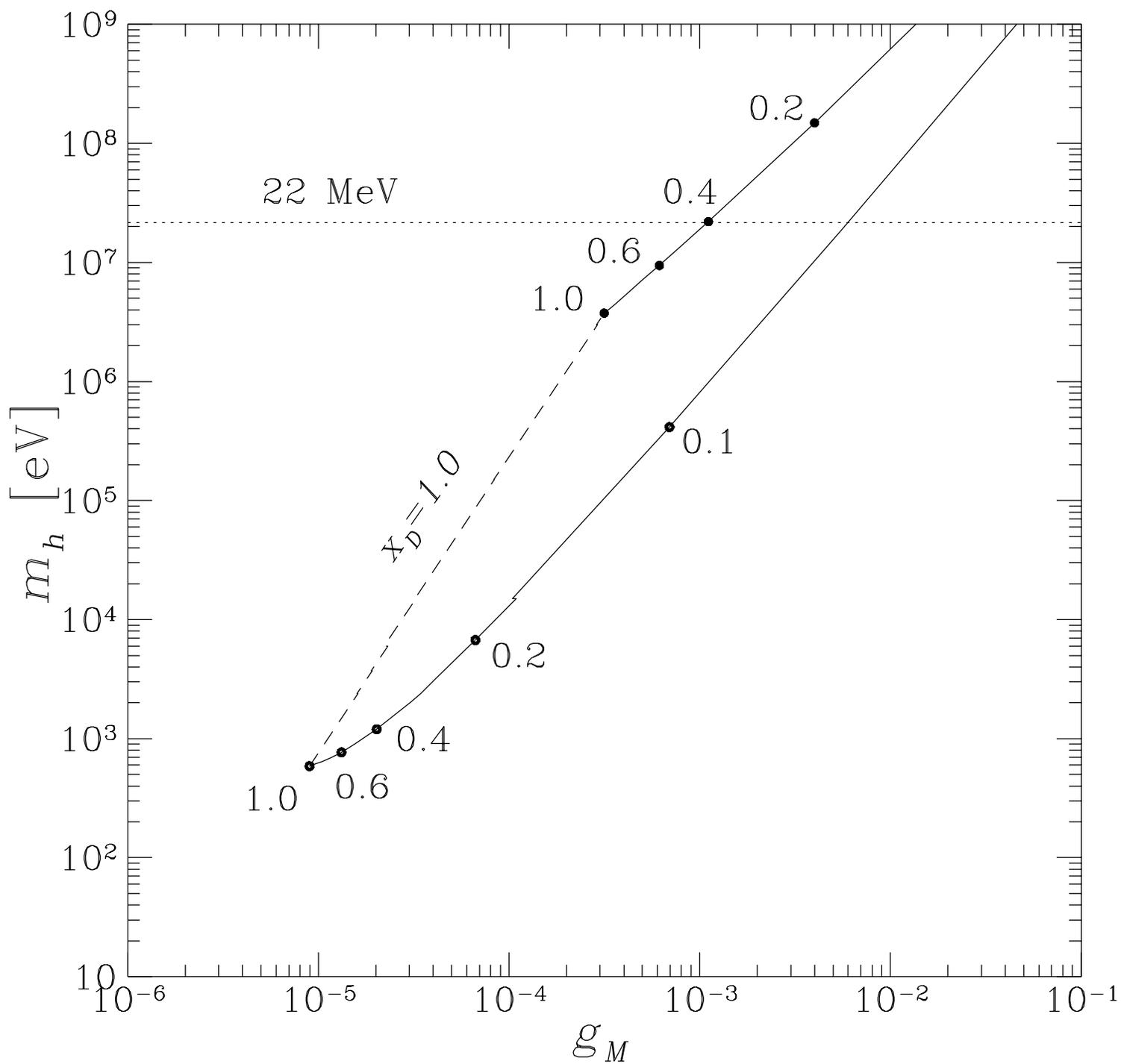
\end{document}